\documentclass{webofc}
\usepackage[varg]{txfonts}   
\usepackage{amsmath}
\begin{document}
\title{The dijet mass distribution in heavy ion collisions}
%
%

\author{\firstname{Jared} \lastname{Reiten}\inst{1,2,3}\fnsep\thanks{\email{jdreiten@physics.ucla.edu}}}

\institute{Department of Physics and Astronomy, University of California, Los Angeles, California 90095, USA
\and Mani L. Bhaumik Institute for Theoretical Physics, University of California, Los Angeles, California 90095, USA
\and Theoretical Division, Los Alamos National Laboratory, Los Alamos, New Mexico 87545, USA}

\abstract{
In these proceedings, we review the production of both light and heavy flavor dijets in heavy ion collisions and highlight a promising observable to expose their distinct signatures. We propose the modification of dijet invariant mass distributions in heavy ion collisions as a novel observable that exhibits striking sensitivity to the quark-gluon plasma transport properties and heavy quark mass effects on in-medium parton showers. This observable has the advantage of amplifying the effects of jet quenching in contrast to conventional observables, such as the dijet momentum imbalance shift, which involve cancellations of such effects and, hence, result in less pronounced signals. Predictions are presented for Au+Au collisions at $\sqrt{s_{NN}}$ = 200 GeV to guide the future sPHENIX program at the Relativistic Heavy Ion Collider.
}
\maketitle

\section{Introduction}
\label{intro}
Heavy flavor jets are a new frontier in the hard probes thrust of the ultra-relativistic nuclear collisions program. Back-to-back dijet measurements, in particular, are exciting experimental channels to study the physics of jet production and propagation in a dense QCD medium -- they provide insight into the path length, color charge, and mass dependence of quark and gluon energy loss in the quark-gluon-plasma (QGP) formed by colliding heavy ions. A clear advantage that heavy flavor dijets have over single inclusive heavy flavor jets is that they are predominately initiated by $Q+\bar{Q}$ pairs in the final partonic state \cite{Kang:2018jr, Huang:2015mva}, whereas single jets receive large contributions from prompt gluons -- splitting to heavy quarks only in later stages of the showering process \cite{Huang:2013vaa, Li:2018xuv}. This means that the study of the modification of dijet observables provides direct information about heavy flavor energy loss mechanisms in the medium, while single jet observables reflect those of light partons, as has been experimentally verified by the CMS collaboration \cite{Chatrchyan:2013exa}.

In performing such measurements, experimentalists typically focus on the most energetic jet, the so-called ``leading'' jet, and the second most energetic jet, the ``subleading'' jet, of a given event \cite{Aad:2010bu, Chatrchyan:2011sx, Khachatryan:2015lha, Adamczyk:2016fqm, Sirunyan:2018jju}. This correlated pair then provides information on energy loss mechanisms in the QGP through the enhancement of their asymmetry, namely, the development of an imbalance between their respective momenta. Such imbalances develop as a result of the differing expanses of medium each jet must traverse before reaching detectors.

In this review, we will focus on two dijet observables to be utilized at the future sPHENIX experiment at RHIC: the traditional modification of the dijet imbalance distribution and the newly proposed modification of the dijet mass distribution \cite{Kang:2018jr}.

\section{Dijet production in heavy ion collisions}
\label{sec-1}

We begin by discussing the calculation of the dijet production cross section in heavy ion collisions. To do so, we first generate the dijet production cross section in elementary p+p collisions using Pythia 8 \cite{Sjostrand:2007gs}, utilizing the kinematic cuts to be used by the sPHENIX collaboration \cite{sPHENIX}. To map this to the case of heavy ions, we must take into account nuclear geometry as well as the effects of medium-induced energy loss. The former induces a dependence on the impact parameter $|\mathbf{b}_{\perp}|$ in the cross section, while the latter accounts for the alteration of phase space that comes from radiative \cite{Zakharov:1997uu, Baier:1996kr, Gyulassy:2000fs, Wiedemann:2000za, Wang:2001ifa, Arnold:2002ja, Zhang:2003wk, Dokshitzer:2001zm} and collisional \cite{Braaten:1991we, Wicks:2005gt, Adil:2006ei, Thoma:2008my, Berrehrah:2013mua, Neufeld:2014yaa} energy loss. Altogether we have:
\begin{align}
\frac{d\sigma ^{AA}\left(|\mathbf{b}_{\perp}|\right)}{dp_{1T}dp_{2T}}=&\int d^2\mathbf{s}_{\perp}T_A\left(\mathbf{s}_{\perp}-\frac{\mathbf{b}_{\perp}}{2}\right)T_A\left(\mathbf{s}_{\perp}+\frac{\mathbf{b}_{\perp}}{2}\right)\notag \\
&\times \sum _{q,g}\int _0^1d\epsilon \frac{P^1_{q,g}\left(\epsilon ; \mathbf{s}_{\perp},|\mathbf{b}_{\perp}|\right)}{1-f_{q,g}^{1\, \rm loss}\left(R ; \mathbf{s}_{\perp},|\mathbf{b}_{\perp}|\right)\epsilon}\int _0^1 d\epsilon'\frac{P^2_{q,g}\left(\epsilon' ; \mathbf{s}_{\perp},|\mathbf{b}_{\perp}|\right)}{1-f^{2\, \rm loss}_{q,g}\left(R ; \mathbf{s}_{\perp},|\mathbf{b}_{\perp}|\right)\epsilon'} \notag \\
&\times \frac{d\sigma_{q,g}^{NN}\left(\left.p_{1T}\middle/\left[1-f^{1\, \rm loss}_{q,g}\left(R ; \mathbf{s}_{\perp},|\mathbf{b}_{\perp}|\right)\epsilon\right]\right.,\left.p_{2T}\middle/\left[1-f^{2\, \rm loss}_{q,g}\left(R ; \mathbf{s}_{\perp},|\mathbf{b}_{\perp}|\right)\epsilon'\right]\right.\right)}{dp_{1T}dp_{2T}}\label{xsecAA} \, ,
\end{align}
where the $T_A$ is known as the thickness function of the optical Glauber model \cite{Miller:2007ri}, $P^{i}_{q,g}\left(\epsilon \right)$ is the probability for partons $i=1,2$ to lose energy fraction $\epsilon$ to medium-induced gluon bremsstrahlung, and $f^{i, \mathrm{loss}}_{q,g}$ is the fraction of energy lost outside each jet cone. Energy may be lost to both radiative and collisional effects, and is determined via
\begin{equation}
f_{q,g}^{\rm loss}\left(R;\mathrm{rad+coll}\right)=\left.1-\left(\int_0^Rdr\int ^E_{\omega _{\rm min}}d\omega \frac{dN^g_{q,g}\left(\omega,r\right)}{d\omega dr} \right)\middle/\left(\int_0^{R_{\rm max}}dr\int ^E_0d\omega \frac{dN^g_{q,g}\left(\omega,r\right)}{d\omega dr} \right)\right. \, ,
\end{equation}
where $\omega _{\rm min}$ parameterizes collisional energy loss.

\begin{figure*}
\centering
\includegraphics[width=6.0cm,clip]{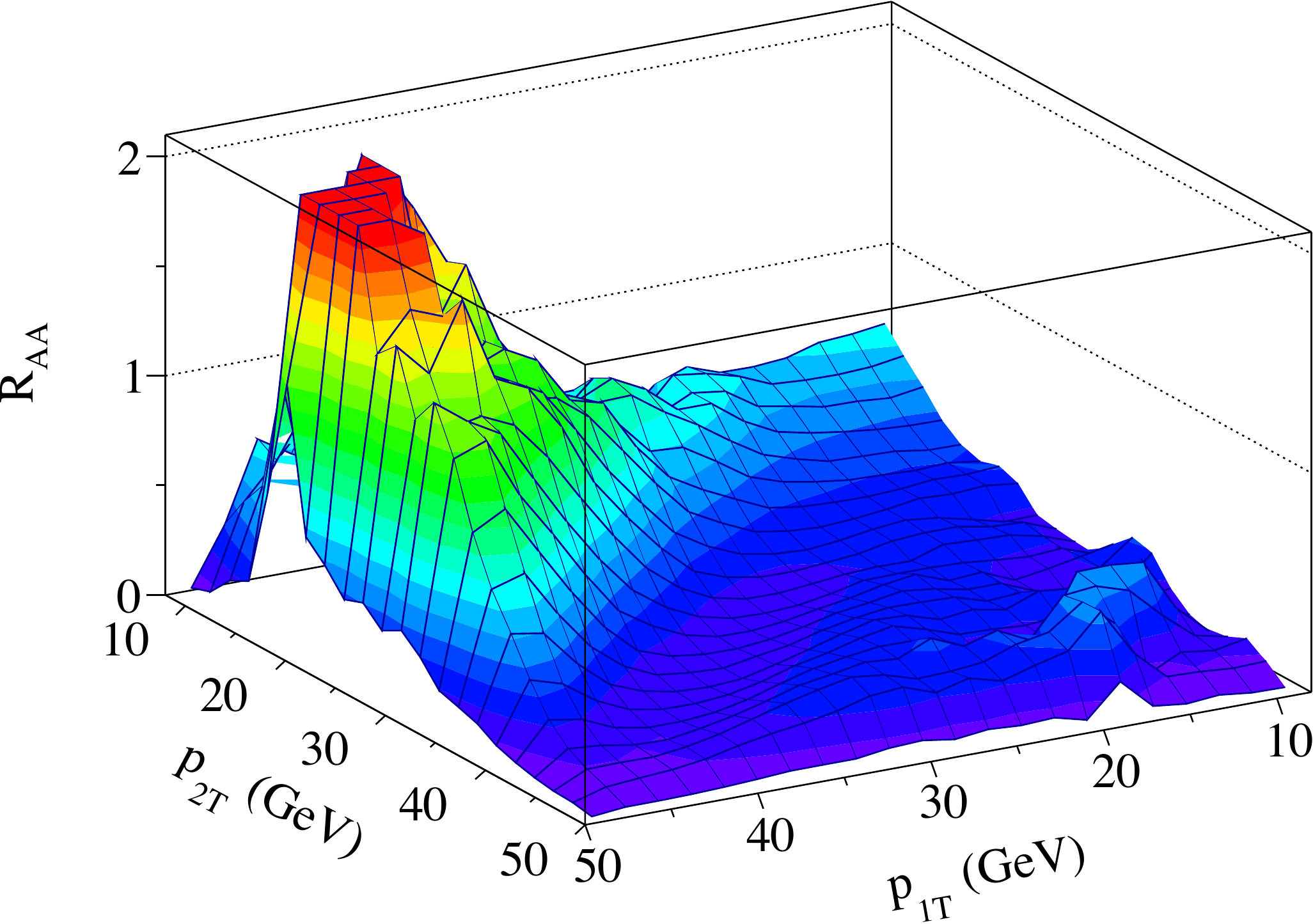}
\hskip 0.1in
\includegraphics[width=6.0cm,clip]{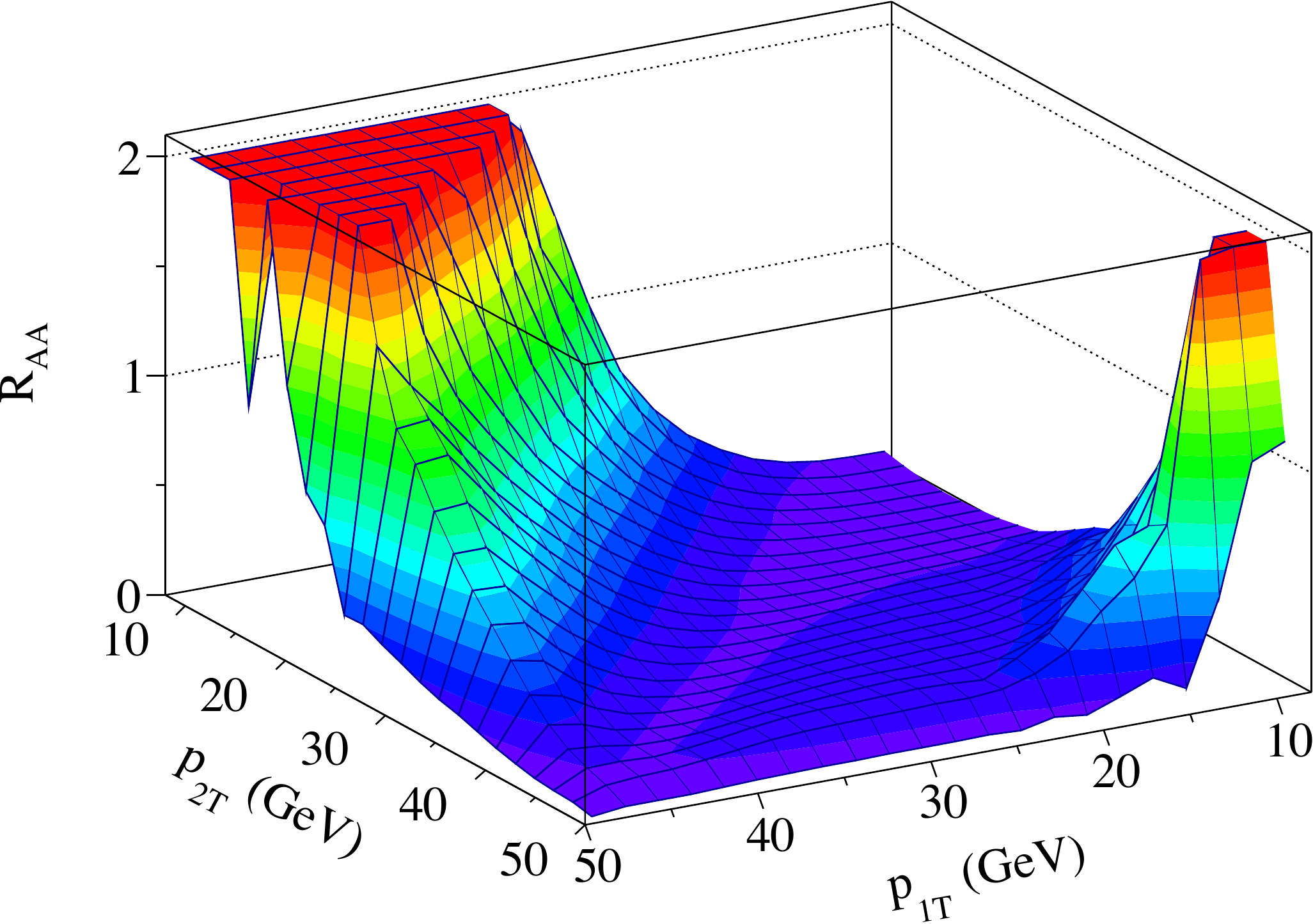}
\caption{Nuclear modification factor for b-tagged (left) and inclusive (right) dijet production in Au+Au collisions at $\sqrt{s_{NN}}=200$ GeV with sPHENIX kinematics \cite{sPHENIX}.}
\label{fig-1}
\end{figure*}

Once we have Eq.~(\ref{xsecAA}), we may then calculate the so-called nuclear modification factor through
\begin{equation}
R_{AA}\left(p_{1T},p_{2T},|\mathbf{b}_{\perp}|\right)=\frac{1}{\langle N_{\rm bin}\rangle}\frac{d\sigma ^{AA}\left(|\mathbf{b}_{\perp}|\right)/dp_{1T}dp_{2T}}{d\sigma ^{pp}/dp_{1T}dp_{2T}} \, ,
\end{equation}
where, in order to make a meaningful comparison to the cross section obtained in the p+p baseline, we have normalized by the average number of binary collisions for the given centrality, $\langle N_{\rm bin} \rangle$. The $R_{AA}$ for light and heavy flavor dijets is displayed in Fig.~\ref{fig-1}. Note that $R_{AA}=1$ denotes no modification relative to p+p, while $R_{AA}>1$ signals enhancement, and $R_{AA}<1$ marks suppression. Further note that Fig.~\ref{fig-1} shows the greatest suppression existing along the main diagonal -- a characteristic feature of the aforementioned asymmetry, which develops for dijets produced in heavy ion collisions.

\section{Phenomenological results at RHIC}
\label{sec-2}
The dijet production cross section Eq.~(\ref{xsecAA}), differential in the transverse momenta of both the leading and subleading jets, provides us with the physical information required to construct the dijet imbalance and mass distributions. This allows us to study their respective modifications induced by the medium.

\subsection{Dijet imbalance distribution}
\label{sec-zJ}
The dijet imbalance distribution in heavy ion collisions may be calculated according to:
\begin{equation}
\frac{d\sigma^{AA}}{dz_J}=\int \, dp_{1T}\,dp_{2T}\,\frac{d\sigma^{AA}}{dp_{1T}dp_{2T}}\, \delta\left(z_J-\frac{p_{2T}}{p_{1T}}\right)\label{zJ}\, ,
\end{equation}
where we have defined the dijet imbalance $z_J$ to be the ratio of the subleading to leading jet transverse momenta, making the allowed values for $z_J$ lie between zero and unity. 

\begin{figure*}
\centering
\includegraphics[width=6.0cm,clip]{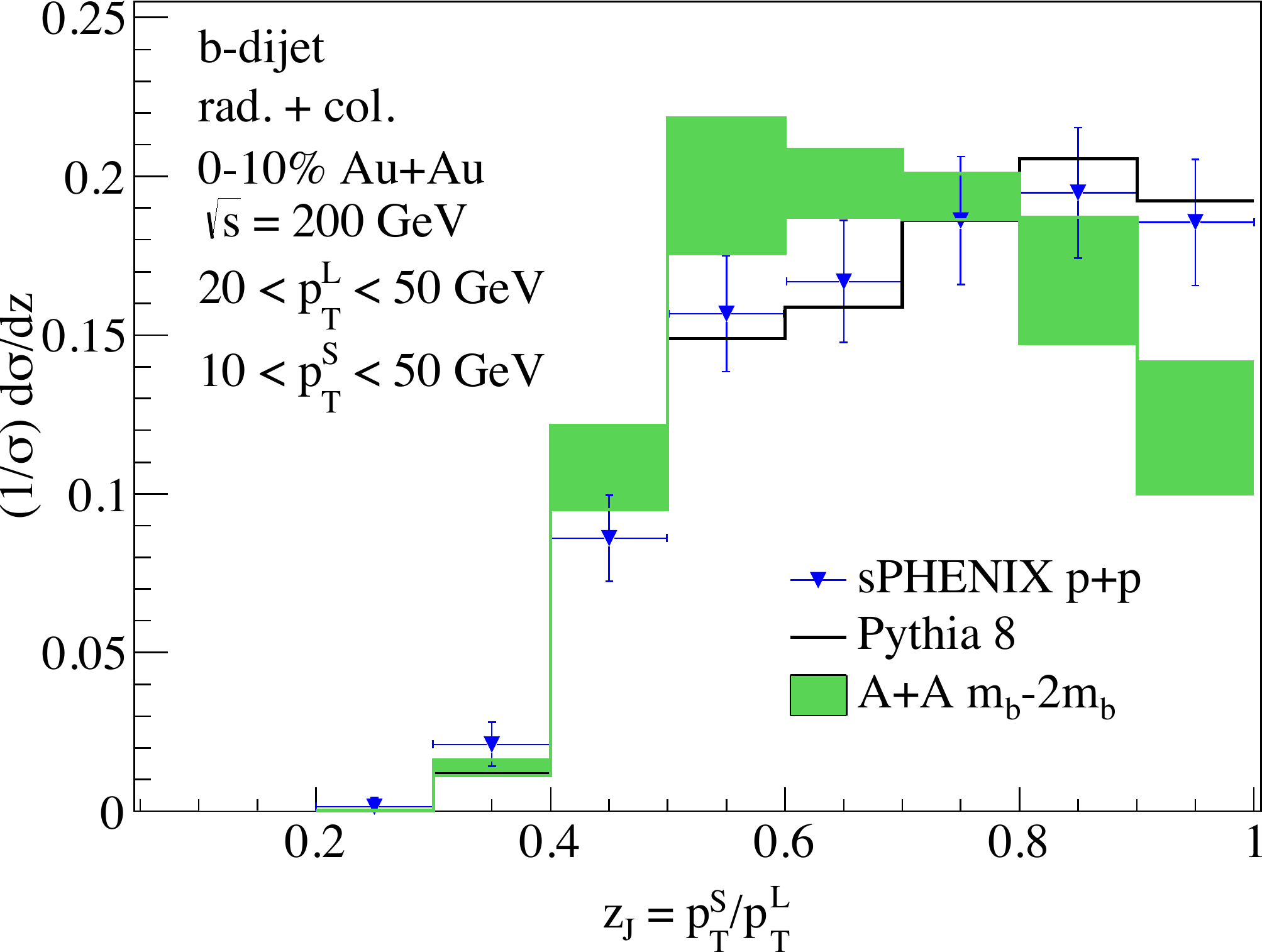}
\hskip 0.1in
\includegraphics[width=6.0cm,clip]{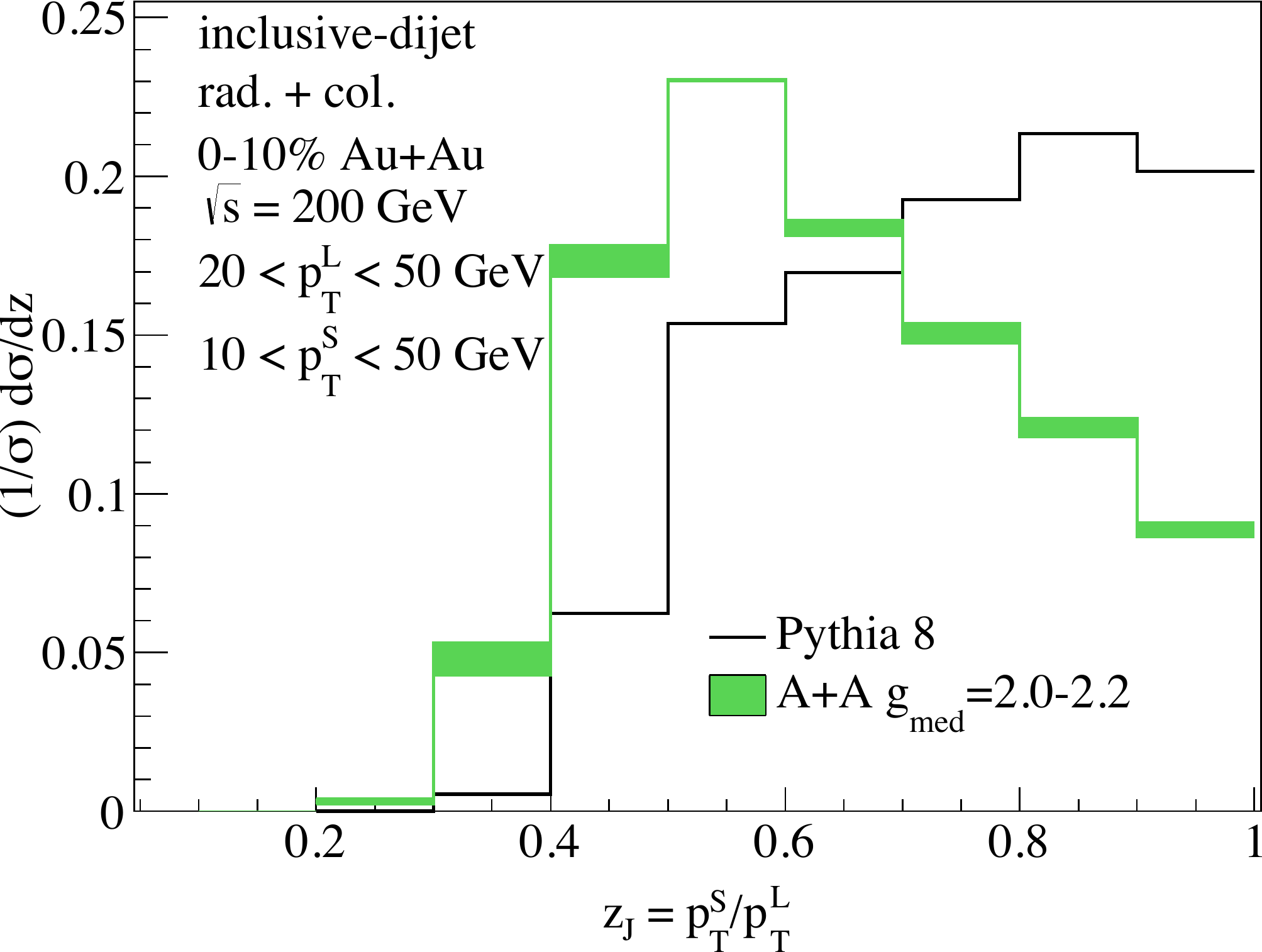}
\caption{The dijet imbalance $z_J$ distributions for b-tagged (left) and inclusive (right) dijet production at $\sqrt{s_{NN}}=200$ GeV with sPHENIX kinematics. The black histogram marks the distribution for p+p collisions, while the green marks that of Au+Au collisions. The blue ``data'' points come from preliminary simulations by the sPHENIX collaboration \cite{sPHENIX}. Left: the coupling of the jet to the medium is fixed at $g_{\rm med}=2.0$ and the band corresponds to varying the range of masses of the propagating system between $m_b$ and $2m_b$. Right: the band corresponds to varying the medium coupling $g_{\rm med}=2.0-2.2$.}
\label{fig-2}
\end{figure*}

Results for this distribution with sPHENIX kinematics are displayed in Fig.~\ref{fig-2}. In this figure, the black histogram is the distribution formed in p+p collisions, while the green is that of A+A. The relative shift between the two is the mark of modification due to the QGP. What is important to note is that the difference between the shift for heavy flavor dijets and that of inclusive (light flavor) is quite subtle. In fact, defining $\Delta \langle z_J \rangle \equiv \Delta \langle z_J \rangle _{pp}-\Delta \langle z_J \rangle _{AA}$, we find that $\Delta \langle z_J\rangle _{bb}=0.065 \pm 0.012$ for heavy flavor dijet and $\Delta \langle z_J\rangle _{jj}=0.100 \pm 0.005$ for light flavor dijet. Such a small effect offers little in differentiation between light and heavy flavors. This leads us to look for other dijet observables that amplify the mass dependence of in-medium parton propagation -- the dijet mass distribution is such an observable.

\subsection{Dijet mass distribution}
\label{sec-m}
The dijet mass distribution in heavy ion collisions can be obtained through use of Eq.~(\ref{xsecAA}) in a fashion analogous to the way Eq.~(\ref{zJ}) is obtained:
\begin{equation}
\frac{d\sigma^{AA}}{dm_{12}}=\int \, dp_{1T}\,dp_{2T}\,\frac{d\sigma^{AA}}{dp_{1T}dp_{2T}} \, \delta\left(m_{12}-\sqrt{\langle m_1^2 \rangle + \langle m_2^2 \rangle + 2p_{1T}\,p_{2T}\langle \mathrm{cosh}\left(\Delta \eta\right) - \mathrm{cos}\left(\Delta \phi\right) \rangle}\right) \label{mass}\, .
\end{equation}
Letting $p_1$ and $p_2$ denote the four-momenta of the jets forming the dijet pair, the dijet invariant mass $m_{12}$ is obtained through $m_{12}=\sqrt{(p_1+p_2)^2}$. The above expression, Eq.~(\ref{mass}), is expressed in the high $p_T$ limit where $p_{iT}\gg m_i$ for $i=1,2$ as well as in terms of the differences in rapidity, $\Delta \eta$, and azimuthal angle, $\Delta \phi$, of the jets forming the dijet pair. Furthermore, the brackets denote averages over each bracketed quantity in p+p collisions, whose modifications in A+A are insignificant, as observed by the ALICE collaboration at the LHC \cite{Acharya:2017goa}.

\begin{figure*}
\centering
\includegraphics[width=6.0cm,clip]{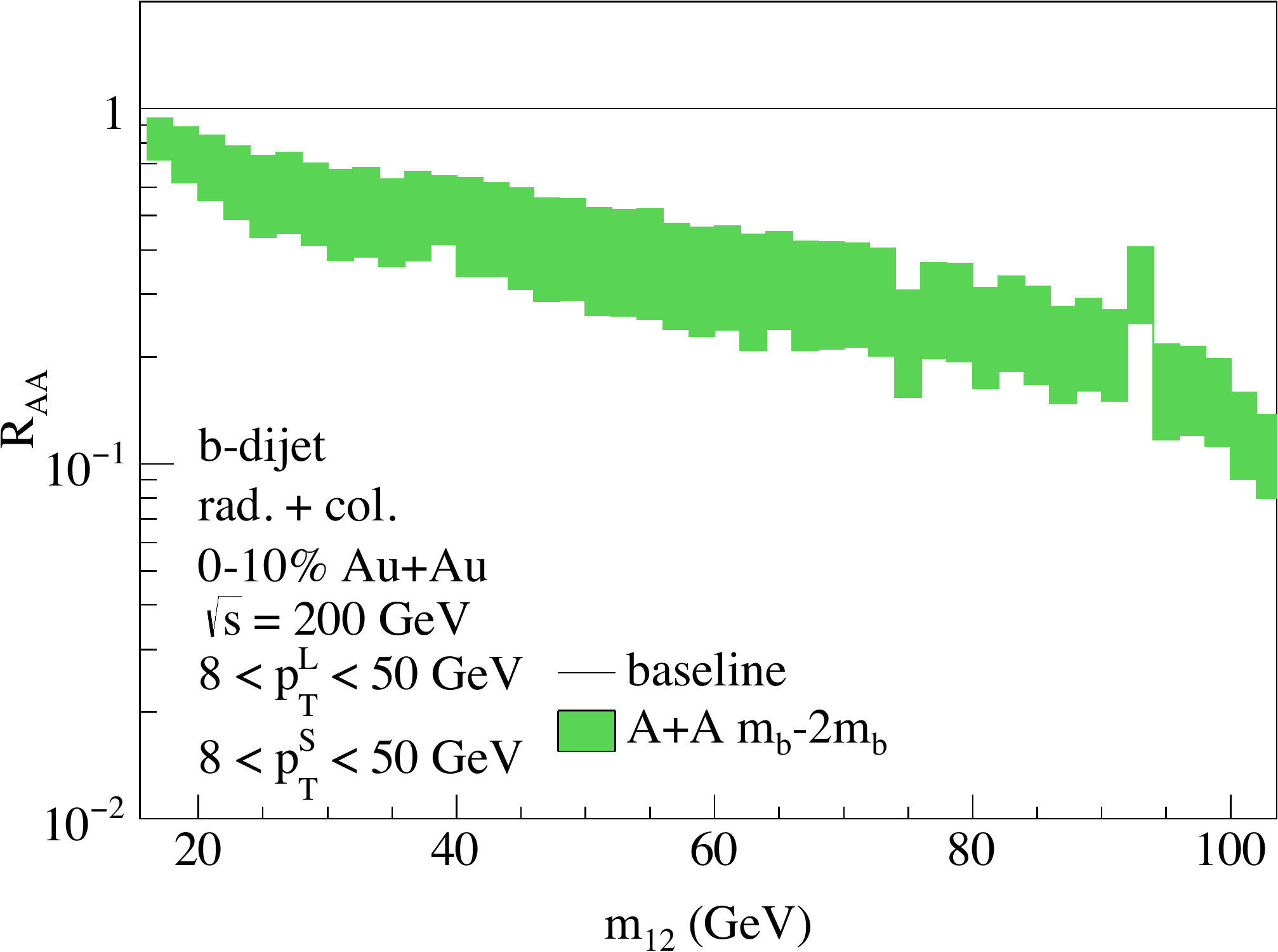}
\hskip 0.1in
\includegraphics[width=6.0cm,clip]{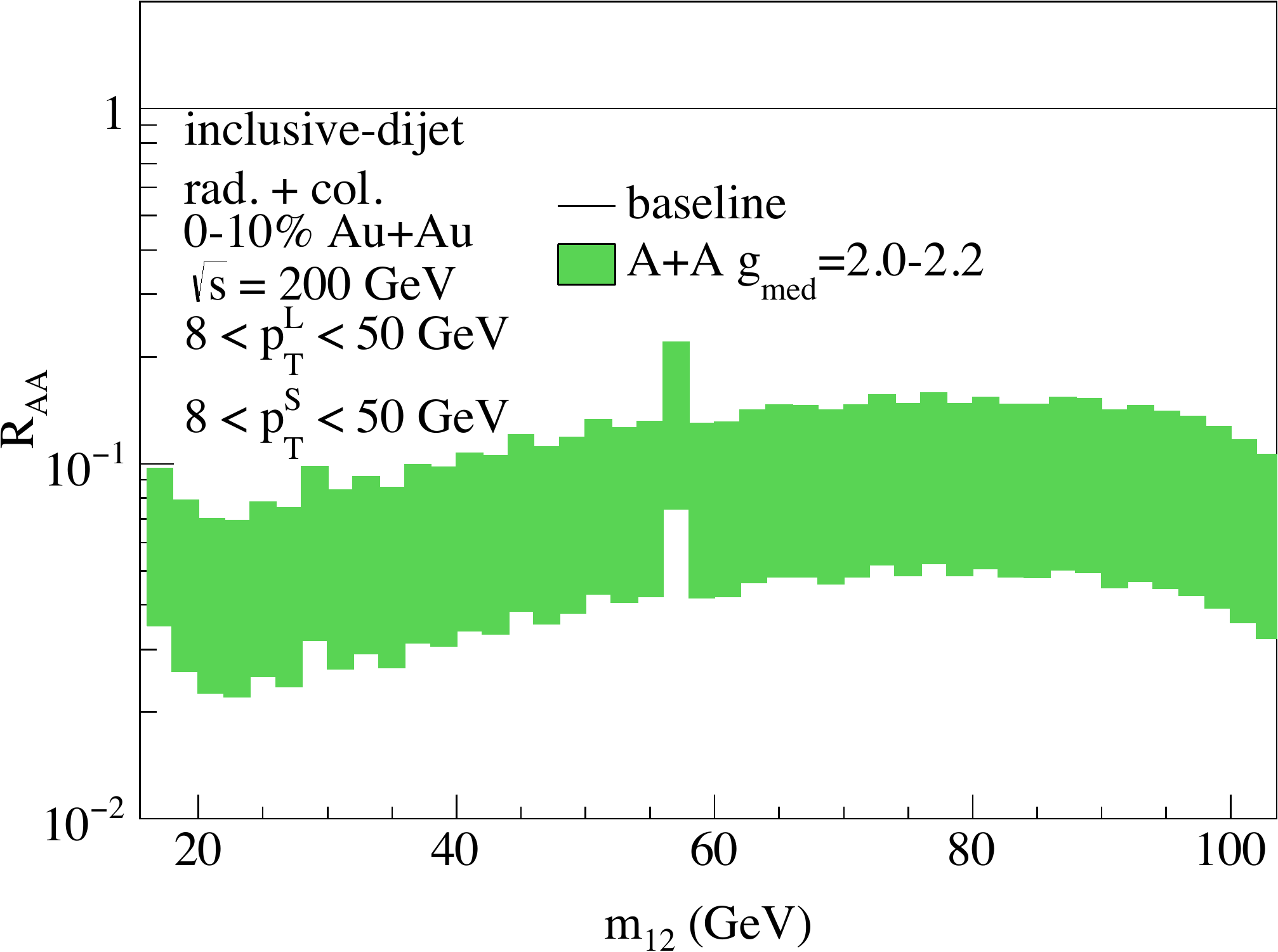}
\caption{Nuclear modification factor $R_{AA}$ with respect to dijet invariant mass $m_{12}$ for b-tagged (left) and inclusive (right) dijet production in Au+Au collisions at $\sqrt{s_{NN}}=200$ GeV for sPHENIX at RHIC. Left: we fix $g_{\rm med}=2.0$ and the band corresponds to varying the mass of the propagating system from $m_b$ to $2m_b$. Right: the band corresponds to the range of medium couplings $g_{\rm med}=2.0-2.2$.}
\label{fig-3}
\end{figure*}

With the dijet mass distribution in hand, we may then define a new observable, namely, the nuclear modification factor for the dijet mass distribution:
\begin{equation}
R_{AA}\left(m_{12},|\mathbf{b}_{\perp}|\right)=\frac{1}{\langle N_{\rm bin}\rangle}\frac{d\sigma ^{AA}\left(|\mathbf{b}_{\perp}|\right)/dm_{12}}{d\sigma ^{pp}/dm_{12}}\, .
\end{equation}
Predictions for this new $R_{AA}$, differential in the dijet invariant mass, are displayed in Fig.~\ref{fig-3} -- these are to guide the future sPHENIX experiment.

\begin{figure}
\centering
\includegraphics[width=6.0cm,clip]{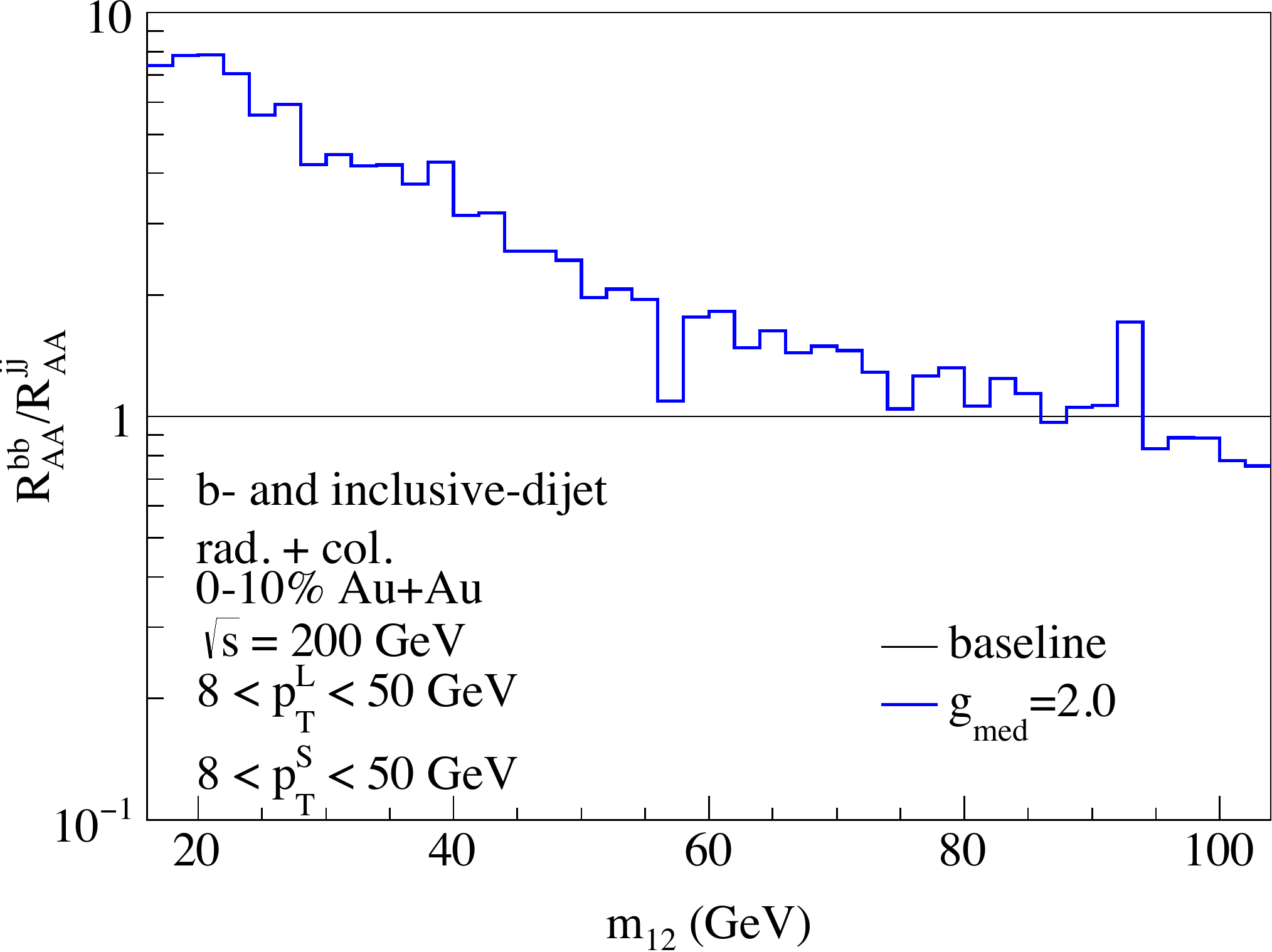}
\caption{Ratio of the nuclear modification factors for b-tagged $\left(R^{bb}_{AA}\right)$ to inclusive $\left(R^{jj}_{AA}\right)$ dijet production vs. dijet invariant mass $m_{12}$ for sPHENIX at RHIC. For b-tagged dijets, the mass of the propagating system is held fixed at $m_b$. For both b-tagged and inclusive dijets, we fix $g_{\rm med}=2.0$.}
\label{fig-4}
\end{figure}

In contrast to the traditional dijet momentum imbalance shift, the dijet mass modification displays a strong and clear dependence on the mass of the partons initiating the dijet. This is exemplified by the factor of $\sim$ 10 or more suppression of inclusive dijets up to a mass of about 100 GeV. Note that this suppression is markedly different from that of b-tagged dijets through most of the mass range covered. This point is highlighted in Fig.~\ref{fig-4}, where we plot the ratio of b-tagged $R_{AA}$ to that of inclusive. Indeed, in the low invariant mass range around 20 GeV, the suppressions experienced by light and heavy flavor dijets differ by almost an order of magnitude. This means that the kinematic range of sPHENIX is particularly well-suited to reveal the mass dependence of quark and gluon energy loss in the QGP -- an important open question in the heavy ion community.

The amplification of quenching effects exhibited in the modification of dijet invariant mass distributions can be understood in the following simple way. Reconstructed jets emerging from A+A collisions have less energy than their counterparts in p+p, hence, in the reverse engineering of A+A cross sections from those of p+p, a jet with a given $p_T$ in A+A is mapped to a jet of a higher value, say $p_T+\delta p_T$, in p+p. This is the qualitative content of Eq.~(\ref{xsecAA}). Now, the dijet imbalance $z_J$ is defined as a ratio of transverse momenta in Eq.~(\ref{zJ}), while the dijet mass $m_{12}$ involves a product of transverse momenta in Eq.~(\ref{mass}). Thus, the effects of the medium between the leading and subleading jets work to cancel one another out in the modification of $z_J$, while they strictly combine to amplify the effects of one another in their modification of $m_{12}$. Therefore, the overall alteration of $z_J$ is more minute, resulting in minimal modification. At the same time, the change to $m_{12}$ is substantial and leads to a sizable plunge down the p+p cross section to obtain that of A+A, resulting in significant modification.

\section{Conclusion}
In these proceedings, we have reviewed the recent predictions for the modification of the dijet mass distribution in heavy ion collisions \cite{Kang:2018jr}. In particular, we have examined the behavior of this observable within the kinematic range of the future sPHENIX experiment at RHIC. The dramatic difference between the modification patterns exhibited by light and heavy flavor dijets makes this observable a promising tool in the growing arsenal of hard probes of the QCD medium. The dijet mass distribution has the advantage of amplifying the effects of jet quenching in heavy ion collisions -- making it stand out among traditional dijet observables, such as the dijet imbalance shift. This is due to its dependence on the product of the transverse momenta of the jets forming the dijet pair, whereas the dijet imbalance distribution depends on their quotient. By amplifying the effects of the medium, the differences in energy loss mechanisms governing the propagation of light and heavy flavor partons within the QGP become more apparent, as displayed in Figs.~\ref{fig-3} and \ref{fig-4}. We look forward to the first experimental measurements of this novel observable, as they are sure to enrich our knowledge of the QGP.

\section{Acknowledgements}
I thank Theoretical Division of Los Alamos National Laboratory for its hospitality during the completion of part of this work. This work is supported by the UC Office of the President through the UC Laboratory Fees Research Program under Grant No. LGF-19-601097. 
%
%

\begin{bibliography}{ismd19.bbl}
\end{bibliography}

\end{document}